\input harvmac
\noblackbox

\input epsf
   
 
\def\journal#1&#2(#3){\unskip, \sl #1\ \bf #2 \rm(19#3) }
\def\andjournal#1&#2(#3){\sl #1~\bf #2 \rm (19#3) }

\def\frac#1#2{{#1\over#2}}

\def\half{\frac12}

\def\d{\partial}

\def\inbar{\,\vrule height1.5ex width.4pt depth0pt}
\def\IC{\relax\hbox{$\inbar\kern-.3em{\rm C}$}}
\def\IR{\relax{\rm I\kern-.18em R}}
\def\IP{\relax{\rm I\kern-.18em P}}
\def\IZ{\relax{\rm I\kern-.18em Z}}
\def\IE{\relax{\rm I\kern-.18em E}}

%
%

%
\catcode`\@=11
\def\slash#1{\mathord{\mathpalette\c@ncel{#1}}}
\overfullrule=0pt

\def\FF{{\cal F}}

\def\II{{\cal I}}

\def\OO{{\cal O}}

\def\QQ{{\cal Q}}

\def\TT{{\cal T}}

\def\ZZ{{\cal Z}}

\def\underrel#1\over#2{\mathrel{\mathop{\kern\z@#1}\limits_{#2}}}

\catcode`\@=12


%

\def \sinh{{\rm sinh}}
\def \cosh{{\rm cosh}}



\def\unlockat{\catcode`\@=11}
\def\lockat{\catcode`\@=12}

\unlockat


\def\newsec#1{\global\advance\secno by1\message{(\the\secno. #1)}
\global\subsecno=0\global\subsubsecno=0\eqnres@t\noindent
{\bf\the\secno. #1}
\writetoca{{\secsym} {#1}}\par\nobreak\medskip\nobreak}
\global\newcount\subsecno \global\subsecno=0
\def\subsec#1{\global\advance\subsecno
by1\message{(\secsym\the\subsecno. #1)}
\ifnum\lastpenalty>9000\else\bigbreak\fi\global\subsubsecno=0
\noindent{\it\secsym\the\subsecno. #1}
\writetoca{\string\quad {\secsym\the\subsecno.} {#1}}
\par\nobreak\medskip\nobreak}
\global\newcount\subsubsecno \global\subsubsecno=0
\def\subsubsec#1{\global\advance\subsubsecno by1
\message{(\secsym\the\subsecno.\the\subsubsecno. #1)}
\ifnum\lastpenalty>9000\else\bigbreak\fi
\noindent\quad{\secsym\the\subsecno.\the\subsubsecno.}{#1}
\writetoca{\string\qquad{\secsym\the\subsecno.\the\subsubsecno.}{#1}}
\par\nobreak\medskip\nobreak}

\def\subsubseclab#1{\DefWarn#1\xdef
#1{\noexpand\hyperref{}{subsubsection}%
{\secsym\the\subsecno.\the\subsubsecno}%
{\secsym\the\subsecno.\the\subsubsecno}}%
\writedef{#1\leftbracket#1}\wrlabeL{#1=#1}}
\lockat


\newcount\figno
\figno=1
\def\fig#1#2#3{
\par\begingroup\parindent=0pt\leftskip=1cm\rightskip=1cm\parindent=0pt
\baselineskip=11pt
\global\advance\figno by 1
\midinsert
\epsfxsize=#3
\centerline{\epsfbox{#2}}
{\bf Fig.\ \the\figno: } #1\par
\endinsert\endgroup\par
}
\def\figlabel#1{\xdef#1{\the\figno}}
\def\encadremath#1{\vbox{\hrule\hbox{\vrule\kern8pt\vbox{\kern8pt
\hbox{$\displaystyle #1$}\kern8pt}
\kern8pt\vrule}\hrule}}
%
%


\font\cmss=cmss10
\font\cmsss=cmss10 at 7pt
\def\rlx{\relax\leavevmode}
\def\inbar{\vrule height1.5ex width.4pt depth0pt}
\def\IC{\relax\,\hbox{$\inbar\kern-.3em{\rm C}$}}
\def\IN{\relax{\rm I\kern-.18em N}}
\def\IP{\relax{\rm I\kern-.18em P}}
\def\ZZ{\rlx\leavevmode\ifmmode\mathchoice{\hbox{\cmss Z\kern-.4em Z}}
 {\hbox{\cmss Z\kern-.4em Z}}{\lower.9pt\hbox{\cmsss Z\kern-.36em Z}}
 {\lower1.2pt\hbox{\cmsss Z\kern-.36em Z}}\else{\cmss Z\kern-.4em
 Z}\fi}
\def\IZ{\relax\ifmmode\mathchoice
{\hbox{\cmss Z\kern-.4em Z}}{\hbox{\cmss Z\kern-.4em Z}}
{\lower.9pt\hbox{\cmsss Z\kern-.4em Z}}
{\lower1.2pt\hbox{\cmsss Z\kern-.4em Z}}\else{\cmss Z\kern-.4em
Z}\fi}
\def\IZ{\relax\ifmmode\mathchoice
{\hbox{\cmss Z\kern-.4em Z}}{\hbox{\cmss Z\kern-.4em Z}}
{\lower.9pt\hbox{\cmsss Z\kern-.4em Z}}
{\lower1.2pt\hbox{\cmsss Z\kern-.4em Z}}\else{\cmss Z\kern-.4em
Z}\fi}

\def\narrowplus{\kern -.04truein + \kern -.03truein}
\def\narrowminus{- \kern -.04truein}
\def\narrowminussub{\kern -.02truein - \kern -.01truein}

\def\half{{1\over 2}}

\def\IZ{\relax\ifmmode\mathchoice
{\hbox{\cmss Z\kern-.4em Z}}{\hbox{\cmss Z\kern-.4em Z}}
{\lower.9pt\hbox{\cmsss Z\kern-.4em Z}}
{\lower1.2pt\hbox{\cmsss Z\kern-.4em Z}}\else{\cmss Z\kern-.4em
Z}\fi}
\def\IB{\relax{\rm I\kern-.18em B}}
\def\IC{{\relax\hbox{$\inbar\kern-.3em{\rm C}$}}}
\def\ID{\relax{\rm I\kern-.18em D}}
\def\IE{\relax{\rm I\kern-.18em E}}
\def\IF{\relax{\rm I\kern-.18em F}}
\def\IG{\relax\hbox{$\inbar\kern-.3em{\rm G}$}}
\def\IGa{\relax\hbox{${\rm I}\kern-.18em\Gamma$}}
\def\IH{\relax{\rm I\kern-.18em H}}
\def\II{\relax{\rm I\kern-.18em I}}
\def\IK{\relax{\rm I\kern-.18em K}}
\def\IP{\relax{\rm I\kern-.18em P}}

\font\cmss=cmss10 \font\cmsss=cmss10 at 7pt
\def\IR{\relax{\rm I\kern-.18em R}}


%

%
%
\def\eqnn#1{\xdef #1{(\secsym\the\meqno)}\writedef{#1\leftbracket#1}%
\global\advance\meqno by1\wrlabeL#1}
\def\eqna#1{\xdef #1##1{\hbox{$(\secsym\the\meqno##1)$}}
\writedef{#1\numbersign1\leftbracket#1{\numbersign1}}%
\global\advance\meqno by1\wrlabeL{#1$\{\}$}}
\def\eqn#1#2{\xdef #1{(\secsym\the\meqno)}\writedef{#1\leftbracket#1}%
\global\advance\meqno by1$$#2\eqno#1\eqlabeL#1$$}


\def\boxit#1{\vbox{\hrule\hbox{\vrule\kern8pt
\vbox{\hbox{\kern8pt}\hbox{\vbox{#1}}\hbox{\kern8pt}}
\kern8pt\vrule}\hrule}}
\def\mathboxit#1{\vbox{\hrule\hbox{\vrule\kern5pt\vbox{\kern5pt
\hbox{$\displaystyle #1$}\kern5pt}\kern5pt\vrule}\hrule}}


\lref\EmparanHG{
  R.~Emparan, T.~Harmark, V.~Niarchos and N.~A.~Obers,
  ``Blackfolds in Supergravity and String Theory,''
JHEP {\bf 1108}, 154 (2011).
[arXiv:1106.4428 [hep-th]].
}

\lref\EmparanAT{
  R.~Emparan, T.~Harmark, V.~Niarchos and N.~A.~Obers,
  ``Essentials of Blackfold Dynamics,''
  JHEP {\bf 1003} (2010) 063
  [arXiv:0910.1601 [hep-th]].
}
\lref\CaldarelliXZ{
  M.~M.~Caldarelli, R.~Emparan and B.~Van Pol,
  ``Higher-dimensional Rotating Charged Black Holes,''
JHEP {\bf 1104}, 013 (2011).
[arXiv:1012.4517 [hep-th]].
}

\lref\HarmarkFF{
  T.~Harmark,
  ``Open branes in space-time noncommutative little string theory,''
Nucl.\ Phys.\ B {\bf 593}, 76 (2001).
[hep-th/0007147].
}

\lref\HarmarkRB{
  T.~Harmark and N.~A.~Obers,
  ``Phase structure of noncommutative field theories and spinning brane bound states,''
JHEP {\bf 0003}, 024 (2000).
[hep-th/9911169].
}

\lref\GrignaniXM{
  G.~Grignani, T.~Harmark, A.~Marini, N.~A.~Obers and M.~Orselli,
  ``Heating up the BIon,''
JHEP {\bf 1106}, 058 (2011).
[arXiv:1012.1494 [hep-th]].
}

\lref\CallanKZ{
  C.~G.~Callan and J.~M.~Maldacena,
  ``Brane death and dynamics from the Born-Infeld action,''
Nucl.\ Phys.\ B {\bf 513}, 198 (1998).
[hep-th/9708147].
}

\lref\FayyazuddinQK{
  A.~Fayyazuddin, T.~Z.~Husain and D.~P.~Jatkar,
  ``One dimensional M5-brane intersections,''
Phys.\ Rev.\ D {\bf 71}, 106003 (2005).
[hep-th/0407129].
}

\lref\BasuED{
  A.~Basu and J.~A.~Harvey,
  ``The M2-M5 brane system and a generalized Nahm's equation,''
Nucl.\ Phys.\ B {\bf 713}, 136 (2005).
[hep-th/0412310].
}

\lref\BermanBV{
  D.~S.~Berman,
  ``M-theory branes and their interactions,''
Phys.\ Rept.\  {\bf 456}, 89 (2008).
[arXiv:0710.1707 [hep-th]].
}

\lref\HoweUE{
  P.~S.~Howe, N.~D.~Lambert and P.~C.~West,
  ``The Self-dual string soliton,''
Nucl.\ Phys.\ B {\bf 515}, 203 (1998).
[hep-th/9709014].
}

\lref\LuninMJ{
  O.~Lunin,
  ``Strings ending on branes from supergravity,''
JHEP {\bf 0709}, 093 (2007).
[arXiv:0706.3396 [hep-th]].
}

\lref\GauntlettCV{
  J.~P.~Gauntlett,
  ``Intersecting branes,''
In *Seoul/Sokcho 1997, Dualities in gauge and string theories* 146-193.
[hep-th/9705011].
}

\lref\SmithWN{
  D.~J.~Smith,
  ``Intersecting brane solutions in string and M theory,''
Class.\ Quant.\ Grav.\  {\bf 20}, R233 (2003).
[hep-th/0210157].
}

\lref\YoumZS{
  D.~Youm,
  ``Partially localized intersecting BPS branes,'' 
Nucl.\ Phys.\ {\bf 556}, 222 (1999). 
[hep-th/9902208].
}

\lref\ChenTT{
  B.~Chen,
  ``The Self-dual String Soliton in $AdS_4\times S^7$ spacetime,''
Eur.\ Phys.\ J.\ C {\bf 54}, 489 (2008).
[arXiv:0710.2593 [hep-th]].
}

\lref\ChenIR{
  B.~Chen, W.~He, J.~-B.~Wu and L.~Zhang,
  ``M5-branes and Wilson Surfaces,''
JHEP {\bf 0708}, 067 (2007).
[arXiv:0707.3978 [hep-th]].
}

\lref\GauntlettAW{
  J.~P.~Gauntlett,
  ``Membranes on five-branes,''
Adv.\ Theor.\ Math.\ Phys.\  {\bf 3}, 775 (1999).
[hep-th/9906162].
}

\lref\CampsHW{
  J.~Camps and R.~Emparan,
  ``Derivation of the blackfold effective theory,''
JHEP {\bf 1203}, 038 (2012).
[arXiv:1201.3506 [hep-th]].
}

\lref\second{
  V.~Niarchos and K.~Siampos,
  to appear.
}  

\lref\GrignaniMR{
  G.~Grignani, T.~Harmark, A.~Marini, N.~A.~Obers and M.~Orselli,
  ``Thermodynamics of the hot BIon,''
Nucl.\ Phys.\ B {\bf 851}, 462 (2011).
[arXiv:1101.1297 [hep-th]].
}

\lref\EmparanCS{
  R.~Emparan, T.~Harmark, V.~Niarchos and N.~A.~Obers,
  ``World-Volume Effective Theory for Higher-Dimensional Black Holes,''
Phys.\ Rev.\ Lett.\  {\bf 102}, 191301 (2009).
[arXiv:0902.0427 [hep-th]].
}

\lref\BhattacharyyaJC{
  S.~Bhattacharyya, V.~E.~Hubeny, S.~Minwalla and M.~Rangamani,
  ``Nonlinear Fluid Dynamics from Gravity,''
JHEP {\bf 0802}, 045 (2008).
[arXiv:0712.2456 [hep-th]].
}

\lref\HubenyHD{
  V.~E.~Hubeny, S.~Minwalla and M.~Rangamani,
  ``The fluid/gravity correspondence,''
13th chapter of Black Holes in Higher Dimensions  (editor: G. Horowitz), Cambridge University Press.
[arXiv:1107.5780 [hep-th]].
}

\lref\RussoIF{
  J.~G.~Russo and A.~A.~Tseytlin,
  ``Waves, boosted branes and BPS states in M-theory,''
  Nucl.\ Phys.\  B {\bf 490}, 121 (1997)
  [arXiv:hep-th/9611047].
}

\lref\ArmasHZ{
   J.~Armas and N.~A.~Obers,
   ``Blackfolds in (Anti)-de Sitter Backgrounds,''
   Phys.\ Rev.\ D {\bf 83}, 084039 (2011).
[arXiv:1012.5081 [hep-th]].
}

\lref\CaldarelliPZ{
   M.~M.~Caldarelli, R.~Emparan and M.J.~Rodriguez,
   ``Black Rings in (Anti)-deSitter space,''
JHEP {\bf 0811}, 011 (2008).
[arXiv:0806.1954 [hep-th]].
}      

\lref\HoweFB{
  P.~S.~Howe, E.~Sezgin and P.~C.~West,
  ``Covariant field equations of the M theory five-brane,''
Phys.\ Lett.\ B {\bf 399}, 49 (1997).
[hep-th/9702008].
}

\lref\BandosUI{
  I.~A.~Bandos, K.~Lechner, A.~Nurmagambetov, P.~Pasti, D.~P.~Sorokin and M.~Tonin,
  ``Covariant action for the superfive-brane of M theory,''
Phys.\ Rev.\ Lett.\  {\bf 78}, 4332 (1997).
[hep-th/9701149].
}

\lref\AganagicZQ{
  M.~Aganagic, J.~Park, C.~Popescu and J.~H.~Schwarz,
  ``World volume action of the M theory five-brane,''
Nucl.\ Phys.\ B {\bf 496}, 191 (1997).
[hep-th/9701166].
}

\lref\GrignaniIW{
  G.~Grignani, T.~Harmark, A.~Marini, N.~A.~Obers and M.~Orselli,
  ``Thermal string probes in AdS and finite temperature Wilson loops,''
[arXiv:1201.4862 [hep-th]].
}

\lref\ArmasUF{
  J.~Armas, J.~Camps, T.~Harmark and N.~A.~Obers,
  ``The Young Modulus of Black Strings and the Fine Structure of Blackfolds,''
JHEP {\bf 1202}, 110 (2012).
[arXiv:1110.4835 [hep-th]].
}

\lref\CampsBR{
  J.~Camps, R.~Emparan and N.~Haddad,
  ``Black Brane Viscosity and the Gregory-Laflamme Instability,''
JHEP {\bf 1005}, 042 (2010).
[arXiv:1003.3636 [hep-th]].
}

\lref\CostaRE{
  M.~S.~Costa,
  ``Black composite M-branes,''
Nucl.\ Phys.\ B {\bf 495}, 195 (1997).
[hep-th/9610138].
}

\lref\IzquierdoMS{
  J.~M.~Izquierdo, N.~D.~Lambert, G.~Papadopoulos and P.~K.~Townsend,
  ``Dyonic membranes,''
Nucl.\ Phys.\ B {\bf 460}, 560 (1996).
[hep-th/9508177].
}

\lref\EmparanUX{
  R.~Emparan, D.~Mateos and P.~K.~Townsend,
  ``Supergravity supertubes,''
JHEP {\bf 0107}, 011 (2001).
[hep-th/0106012].
}



\rightline{CCTP-2012-05}
\rightline{CPHT-RR019.0412}
\vskip 1pt
\Title{
}
{\vbox{\centerline{M2-M5 blackfold funnels}
}}
\medskip
\centerline{Vasilis Niarchos\footnote{$^\flat$}{niarchos@physics.uoc.gr, 
$^\natural$ ksiampos@cpht.polytechnique.fr} and Konstadinos Siampos$^\natural$}
\bigskip
\centerline{{\it $^\flat$Crete Center for Theoretical Physics}}
\centerline{\it Department of Physics, University of Crete, 71303, Greece}
\bigskip
\centerline{{\it $^\natural$Centre de Physique Th\'eorique, \'Ecole Polytechnique}}
\centerline{\it CNRS-UMR 7644, 91128 Palaiseau Cedex, France}
\bigskip\bigskip\bigskip
\centerline{\bf Abstract}
\bigskip

\noindent
We analyze the basic M2-M5 intersection in the supergravity regime using the blackfold
approach. This approach allows us to recover the 1/4-BPS self-dual string soliton solution 
of Howe, Lambert and West as a three-funnel solution of an effective fivebrane worldvolume
theory in a new regime, the regime of a large number of M2 and M5 branes. In addition,
it allows us to discuss finite temperature effects for non-extremal self-dual string soliton solutions 
and wormhole solutions interpolating between stacks of M5 and anti-M5 branes. The purpose
of this paper is to exhibit these solutions and their basic properties.

\vfill
\Date{}



\newsec{M2-M5}
\seclab\intro

The star of this short note is the M2-M5 intersection 
\eqn\introaa{\eqalign{
\vbox{ \offinterlineskip  \halign
{ # & #  & #  &  #  & #  &  # &  #  &  #  & #  & # & # &  #  \cr
      &  0 & 1  & 2   & 3  & 4  & 5   & 6   & 7  & 8 & 9 & 10 \cr 
    \strut &&&&&&&&&&&\cr
M2~:~  & $\bullet$ & $\bullet$ & & & &  & $\bullet$ &&&&    \cr
  \strut &&&&&&&&&&&\cr
M5~:~  & $\bullet$  & $\bullet$ & $\bullet$ & $\bullet$ & $\bullet$ & $\bullet$ & & & & &   \cr}}
      }}
in $\IR^{1,10}$. This setup, which is 1/4-BPS at extremality, is interesting for a variety of
reasons. The (1+1)-dimensional intersection is a self-dual string whose properties underlie many of
the mysteries of the M2 and M5 brane physics and M-theory itself (for a relevant review we refer
the reader to \BermanBV). In the past it has been studied from several points of view:
\item{(1)} As a supersymmetric soliton solution of the effective fivebrane worldvolume theory of a 
single M5 brane \HoweUE. The solution, which preserves the requisite $SO(1,1)\times SO(4) \times 
SO(4)$ symmetry, has a non-trivial worldvolume self-dual three-form flux and a single non-trivial 
transverse scalar field $z:=x^6$ with the profile
\eqn\introab{
z(\sigma)=\frac{2Q_{sd}}{\sigma^2}
~.}
$\sigma$ denotes the radial distance in the directions (2345) transverse to the self-dual string along
the fivebrane worldvolume. $Q_{sd}$ is the electric (equally magnetic) charge of the self-dual string.
\item{(2)} As a three-funnel solution of the Basu-Harvey equation \BasuED, which has been proposed
as an M-theoretic generalization of the Nahm equations for the BIon. This is an alternative M2-based
description of (1) that refers again to the case of a single M5 brane.
\item{(3)} As a 1/4-BPS supergravity solution in the regime of a large number of M2 and M5 branes. 
There has been considerable work in this direction (see \refs{\GauntlettCV\SmithWN-\YoumZS} for 
earlier studies). A fully localized intersection was described in \LuninMJ, where the solution is given 
in terms of two functions that obey a set of differential equations.

In what follows we will describe the M2-M5 intersection from the supergravity perspective $(3)$ 
using the blackfold formalism \refs{\EmparanCS\EmparanAT-\EmparanHG}. This is an effective 
worldvolume description of black brane dynamics which is part of a perturbative expansion scheme 
of the gravitational equations and belongs conceptually to the same class of ideas as the fluid-gravity 
correspondence for AdS black branes \refs{\BhattacharyyaJC,\HubenyHD}. In the present case we 
will be interested in the effective fivebrane worldvolume dynamics of the M2-M5 bound state 
\refs{\IzquierdoMS\CostaRE\RussoIF\HarmarkRB-\HarmarkFF}. We will restrict our attention to the 
leading order form of this effective description assuming small accelerations in a derivative 
expansion in a manner very similar in spirit to the zero-th order approximation typically employed in 
applications of the Dirac-Born-Infeld action for standard D-branes in string theory. As in the case of 
the BIon solution for the F1-D3 system \CallanKZ, we will see that the zero-th order solution can take 
us far enough.

Although a perturbative reconstruction of the exact supergravity solution is in principle possible in 
this manner, working with an effective worldvolume description ---as we will do here--- has some 
obvious benefits. First, we get very quick access to the supersymmetric self-dual string soliton
solution in a clear intuitive manner that extends the single M5 brane worldvolume description 
of \HoweUE\ into the regime of a large number of M5 branes. A non-trivial profile for a single 
transverse scalar analogous to \introab\ is immediately derivable
and the scalar charge is determined in terms of the number of both the M2 and M5 branes.
Second, with very little extra effort we get immediate access to non-extremal configurations 
describing black M2-M5 intersections, which are currently beyond the technical capability of the exact 
solution generating techniques of Ref.\ \LuninMJ. The leading order thermodynamic properties of
these solutions can be determined straightforwardly without the need to refer directly to 
the full supergravity background. A novel treatment of the self-dual string soliton in previously 
inaccessible regimes becomes possible in this way.

We should note that a similar treatment of the BIon solution for the F1-D3 system was given
recently in two beautiful papers \refs{\GrignaniXM,\GrignaniMR}. The F1-D3 intersection is U-dual to 
the M2-M5 system \introaa\ and inevitably the application of the blackfold formalism in 
\refs{\GrignaniXM,\GrignaniMR} shares several common features with the application in this 
note.\foot{Recall, however, that the direct application of U-dualities in supergravity typically does not 
produce fully localized intersections. Regarding the specific relation between the F1-D3 and M2-M5 
systems this point is also noted, and properly taken into account, in \LuninMJ.} Our goal is to highlight 
those features that are particularly interesting from an M-theory perspective and contrast them with 
the results in the existing literature as a basis for further work in this direction. 

The basics of the blackfold approach and the elements we need for the present application are
summarized in section 2. Section 3 presents the main results of this paper, which include the 
self-dual string soliton solution and the key properties of related wormhole solutions. A more detailed 
and more general treatment of the system will be given in a companion paper \second. A brief 
discussion of the results and future directions appears in the concluding section 4.

\newsec{M2-M5 blackfold equations}
\seclab\equations

\subsec{Planar M2-M5 bound state}

Our starting point is a well-known exact solution of the eleven dimensional supergravity equations of 
motion that describes the black M2-M5 bound state 
\refs{\IzquierdoMS\CostaRE\RussoIF\HarmarkRB-\HarmarkFF}
\eqn\planaraa{\eqalign{
ds_{11}^2=&(HD)^{-1/3}\Big[ -f dt^2+(dx^1)^2+(dx^2)^2+D\left( (dx^3)^2+(dx^4)^2+(dx^5)^2 \right)
\cr
&+H\left( f^{-1} dr^2 + r^2 d\Omega_4^2 \right) \Big]
~,}}
\eqn\planarab{
C_3=-\sin\theta(H^{-1}-1) \coth\alpha \, dt\wedge dx^1 \wedge dx^2
+\tan\theta DH^{-1} dx^3\wedge dx^4\wedge dx^5
~,}
\eqn\planarac{
C_6=\cos\theta D(H^{-1}-1)\coth\alpha \, dt\wedge dx^1\wedge \cdots \wedge dx^5
~,}
\eqn\planarad{
H=1+\frac{r_0^3 \sinh^2\alpha}{r^3}~, ~~
f=1-\frac{r_0^3}{r^3}~, ~~ 
D^{-1}=\cos^2\theta+\sin^2\theta H^{-1}
~.}
$C_3$ is the standard three-form potential of the 11d supergravity action and $C_6$ its Hodge-dual.
The solution, which describes M2 brane charge dissolved into the worldvolume of the black 
fivebrane along the (012) plane, is parameterized by the constants $r_0$, $\alpha$ and $\theta$ 
which control the temperature, the M2 and the M5 brane charge. 

The thermodynamic properties of the solution are captured by the following quantities \HarmarkRB
\eqn\planarae{\eqalign{
\varepsilon=\frac{\Omega_{(4)}}{16\pi G}&r_0^3 (4+3\sinh^2 \alpha)
~,~~
\TT=\frac{3}{4\pi r_0 \cosh \alpha}~,~~
s=\frac{\Omega_{(4)}}{4 G} r_0^4 \cosh\alpha
~,
\cr
Q_5=&\cos\theta \, Q~, ~~ \QQ_2=-\sin\theta \, Q~, ~~
Q=\frac{\Omega_{(4)}}{16\pi G} 3 r_0^3 \sinh\alpha\, \cosh\alpha
~,
\cr
&\Phi_5=\cos\theta \, \Phi~, ~~ \Phi_2=-\sin\theta \, \Phi~, ~~
\Phi=\tanh\alpha
~.}}
$\varepsilon$ denotes the energy density, $\TT$ the temperature, $s$ the entropy density,
$Q_5$ the fivebrane charge, $\QQ_2$ the twobrane charge density, and $\Phi_5, \Phi_2$
the corresponding chemical potentials. We will reserve the notation $Q$ for charges and
$\QQ$ for charge densities. In this notation $\QQ_5=Q_5$. A corresponding free energy 
$\FF$ can be defined as
\eqn\planaraj{
\FF=\varepsilon-\TT s=\frac{\Omega_{(4)}}{16\pi G}r_0^3 (1+3\sinh^2 \alpha)
~,}
where $\Omega_{(n)}$ denotes the volume of the round $n$-sphere.

Under the general boost along the fivebrane worldvolume directions
the stress-energy tensor of the above solution takes the form
\eqn\planarak{
T_{ab}=\TT s\, \left( u_a u_b-\frac{1}{3} \eta_{ab} \right) -\sum_{q=2,5} \Phi_q \QQ_q \, h_{ab}^{(q)}
~, ~~ a,b,...=0,1,\ldots,5
}
where, following closely the notation of \EmparanHG, $u^a$ denotes a unit 6-velocity field,
$\eta_{ab}$ the flat induced worldvolume metric, and $h^{(q)}_{ab}$ $(q=2,5)$ is a projector along 
the worldvolume directions of the M2 and M5 branes respectively. In the case of 
\planaraa -\planarad\ $h^{(2)}_{ab}$ projects along the plane $(012)$ and $h^{(5)}_{ab}=\eta_{ab}$.

\subsec{Leading order blackfold equations}

One can use the planar solution \planaraa-\planarad\ as the zero-th order term in a 
perturbative expansion to construct more complicated solutions with inhomogeneous, spinning,
and bending worldvolume geometries. The effective degrees of freedom in such a long-wavelength 
description are the parameters $r_0$, $\alpha$, $\theta$, and the velocities $u^a$ that characterize 
the zero-th order solution, five transverse scalars that capture the bending of the M5 in its 
eleven dimensional ambient space, and a unit three-form that captures the local M2 brane 
current and its distribution within the larger fivebrane worldvolume. The self-dual three-form
field strength of the M5 brane has disappeared in this regime and has been 
replaced by corresponding conserved currents. 

In analogy to usual practice in the fluid-gravity
correspondence for AdS black branes, one promotes the above parameters to slowly varying 
functions of the local worldvolume coordinates $\sigma^a$ $(a=0,1,\ldots,5)$ and 
proceeds to solve the gravitational equations perturbatively in a derivative expansion. A subset 
of the gravity equations are constraint equations. Satisfying them at a given order $n$ is believed to 
guarantee the existence of a regular solution up to the $(n+1)$-th order in the expansion scheme 
(see \CampsHW\ for a recent derivation of this statement for $n=0$ in pure Einstein gravity
and \refs{\CampsBR,\ArmasUF} for a discussion of higher derivative corrections). 

This is the general framework of the blackfold formalism. In what follows we will restrict our attention 
to the leading order constraint equations which are believed to guarantee the existence of 
a regular supergravity solution up to the next-to-leading order in the expansion. In our case 
these equations can be formulated as follows (for more details we refer the reader to 
\refs{\EmparanAT,\CaldarelliXZ,\EmparanHG}).

\medskip
\noindent
{\it Intrinsic equations.}

They comprise of the fluid-dynamical equations
\eqn\eqsca{
D_a T^{ab}=0
~}
and the charge conservation equations
\eqn\eqsab{
d*J_3=0~, ~~ J_3=\QQ_2 \hat V_{(3)}
~,} 
\eqn\eqsabb{
d*J_6=0~, ~~ J_6=\QQ_5 \hat V_{(6)}
}
for the M2 and M5 brane currents respectively. The latter trivially leads to 
\eqn\eqscb{
\d_a Q_5=0
~.}
In these equations
the stress-energy tensor \planarak\ is promoted to 
\eqn\planarak{
T_{ab}=\TT s \left(  u_a u_b-\frac{1}{3} \gamma_{ab} \right)-\sum_{q=2,5} \Phi_q \QQ_q \, h_{ab}^{(q)}
~,}
where $\gamma_{ab}$ is now the general induced worldvolume metric. $\hat V_{(3)}$ denotes a 
unit volume 3-form along the directions of the M2 brane current and $\hat V_{(6)}$ the unit volume 
form of the fivebrane worldvolume. The last equation \eqscb\ shows that $Q_5$ is an overall constant 
that participates passively into the dynamics. The M2 brane charge density and its distribution inside 
the fivebrane worldvolume, however, are dynamical quantities controlled by \eqsca, \eqsab.

\medskip
\noindent
{\it Extrinsic equations.}

These comprise of the remaining set of the stress-energy conservation equations and can be
recast into the form 
\eqn\eqsba{
K_{ab}{}^\rho T^{ab}=0
}
or equivalently in a more detailed form as
\eqn\eqsbb{
\TT s \perp^\rho_{~\mu}\dot u^\mu=\frac{1}{3} \TT s K^\rho
+\perp^\rho_{~\mu} \sum_q \Phi_q \QQ_q K^\mu_{(q)}
~.}
$K_{ab}{}^\rho$ is the extrinsic curvature tensor \EmparanAT, $K^\rho$ the mean curvature
vector, $\perp^\rho_{~\mu}$ a projector in directions orthogonal to the fivebrane worldvolume and 
\eqn\eqsbc{
K^\mu_{(q)}=h^{ab}_{(q)} K_{ab}^{~~\mu}
~.}

In the following section we are looking for simple static solutions of the above equations.

\newsec{Static $S^3$-funnel solutions}
\seclab\solutions

\subsec{Static ansatz}

In what follows we will concentrate on a rather restricted simple class of static $S^3$-funnel solutions
that extend the self-dual string soliton of Ref.\ \HoweUE\ to our context. 
More general solutions are possible and will be discussed in a companion paper along with a
more detailed exposition of the relevant steps. 

We will make use of the following parametrization of the ambient eleven dimensional flat spacetime
\eqn\statioba{
ds^2_{11}=-dt^2+(dx^1)^2+dr^2+r^2 d\Omega_3^2+\sum_{i=6}^{10} (dx^i)^2
}
using the standard angular coordinates $(\psi,\varphi,\omega)$ to express the round three-sphere
metric
\eqn\statiobb{
d\Omega_3^2=d\psi^2+\sin^2\psi(d\varphi^2+\sin^2\varphi \, d\omega^2)
~.}
We choose the static gauge 
\eqn\statiobc{\eqalign{
&t(\sigma^a)=\sigma^0~, ~~ x^1(\sigma^a)=\sigma^1~, ~~ r(\sigma^a)=\sigma^2:=\sigma~,
\cr
&\psi(\sigma^a)=\sigma^3~, ~~ \varphi(\sigma^a)=\sigma^4~, ~~ \omega(\sigma^a)=\sigma^5~, ~~
x^6(\sigma^a)=z(\sigma)
}}
activating only one of the transverse scalars $x^6:=z(\sigma)$ in accordance with \introaa.
With this ansatz the induced metric on the effective fivebrane worldvolume is
\eqn\statiobe{
\gamma_{ab}d\sigma^a d\sigma^b=-(d\sigma^0)^2+(d\sigma^1)^2+(1+{z'}^2)d\sigma^2
+\sigma^2(d\psi^2 +\sin^2\psi(d\varphi^2+\sin^2\varphi d\omega^2))
~.}

By setting 
\eqn\statiobf{
u^a=\left( \frac{\d}{\d t}\right)^a~, ~~~ 
h^{(2)}={\rm diag}(-1,1,1+{z'}^2,0,0,0)
}
and by demanding that the quantities $q_2,q_5,\beta$, defined as
\eqn\statiobl{
q_2:=-\frac{16\pi G}{3\Omega_{(4)}\Omega_{(3)}}Q_2=\sigma^3 \frac{r_0^3}{2}  \sin\theta\, \sinh 2\alpha
~,}
\eqn\statiobm{
q_5:=\frac{16\pi G}{3\Omega_{(4)}}Q_5=\frac{r_0^3}{2} \cos\theta \, \sinh 2\alpha
~,}
\eqn\statiobn{
r_0\,  \cosh \alpha=\beta:=\frac{3}{4\pi T}
}
are constants of motion independent of $\sigma^a$, one can show that the intrinsic equations 
\eqsca, \eqsab, \eqscb\ are fully satisfied. In these relations $Q_2$ and $Q_5$ are the total M2 and 
M5 brane charges, expressed in terms of the number of M2 and M5 branes ($N_2, N_5$) as
\eqn\statiobo{
Q_2=\frac{N_2}{(2\pi)^2 \ell_P^3}
~,~~
Q_5=\frac{N_5}{(2\pi)^5 \ell_P^6}
~.}
$T$ is the global constant temperature of the solution and $\ell_P$ the Planck scale (in terms of
which $16\pi G=(2\pi)^8\ell_P^9$).

These expressions allow us to determine completely the dynamics of the unknown functions 
$r_0, \alpha, \theta$. After a minor algebraic
computation one finds two solutions (both acceptable) with
\eqn\statiocf{
\cosh \alpha_{\pm}=\frac{\beta^3}{\sqrt{2} q_5}
\frac{\sqrt{1\pm \sqrt{1-\frac{4q_5^2}{\beta^6}\left(1+\frac{\kappa^2}{\sigma^6}\right)}}}
{\sqrt{1+\frac{\kappa^2}{\sigma^6}}}
~,}
\eqn\statiocg{
r_{0,\pm}=\frac{\sqrt{2}q_5}{\beta^2}
\frac{\sqrt{1+\frac{\kappa^2}{\sigma^6}}}
{\sqrt{1\pm \sqrt{1-\frac{4q_5^2}{\beta^6}\left(1+\frac{\kappa^2}{\sigma^6}\right)}}}
~,}
\eqn\statiocb{
\tan \theta=\frac{\kappa}{\sigma^3}
~.}
We are using the convenient definition 
\eqn\statioca{
\kappa:=\frac{q_2}{q_5}=-\frac{1}{2\pi^2}\frac{Q_2}{Q_5}=-4\pi \frac{N_2}{N_5} \ell_P^3
~.}
It is also worth noting that the first expression \statiocf\ implies an upper bound on the temperature 
$T$
\eqn\statiocd{
\beta^3 \geq 2 |q_5|
~.}

The final step requires solving the extrinsic equations \eqsba. Inserting the solutions 
\statiocf-\statiocb\ into \eqsba\ we obtain equations of motion exclusively 
for the transverse scalars. It has been shown
\refs{\EmparanAT,\CaldarelliXZ,\EmparanHG} on general grounds for stationary configurations that 
these equations can also be obtained from the variation of the action functional
\eqn\statioae{
I:=\int_{{\cal W}_6} d^6\sigma \sqrt{-\gamma}~ \FF
~,}
where $\FF$ is the free energy \planaraj\ viewed as a functional of the transverse scalars, and
the variation with respect to the transverse scalars is performed keeping the temperature and 
corresponding charges fixed. ${\cal W}_6$ is the six dimensional fivebrane worldvolume. In the case at hand the action \statioae\ becomes
\eqn\statioci{
I=\frac{\Omega_{(3)}\Omega_{(4)} L_t L_{x^1}}{16\pi G} 
\frac{2^{3/2} q_5^3}{\beta^6}
\int d\sigma\, \sqrt{1+{z'}^2}\, F_{\pm}(\sigma)
~,}
\eqn\statiock{
F_\pm (\sigma)=\sigma^3 
\left( \frac{1+\frac{\kappa^2}{\sigma^6}}
{1\pm \sqrt{1-\frac{4q_5^2}{\beta^6}\left(1+\frac{\kappa^2}{\sigma^6}\right)}}\right)^{\frac{3}{2}}
\left( -2+\frac{3\beta^6}{2q_5^2}
\frac{1\pm \sqrt{1-\frac{4q_5^2}{\beta^6}\left(1+\frac{\kappa^2}{\sigma^6}\right)}}
{1+\frac{\kappa^2}{\sigma^6}} \right)
~.}
$L_t$, $L_{x^1}$ denote the (infinite) length of the $t,x^1$ directions. We conclude that 
the corresponding equation of motion for the transverse scalar field $z(\sigma)$ is
\eqn\statiocm{
\left( \frac{z'(\sigma) F_\pm (\sigma)}{\sqrt{1+z'(\sigma)^2}}\right)'=0
~.}
The prime denotes differentiation with respect to $\sigma$.

In analogy to the BIon case \refs{\CallanKZ,\GrignaniXM} we will find spike and wormhole 
solutions to this equation representing M2 branes ending on M5 branes or M2 branes 
stretching between M5 and anti-M5 branes. The leading order approximation is valid
as long as the following condition is met \refs{\GrignaniXM,\second}
\eqn\statioco{
\sigma \gg r_c(\sigma)\ ,\qquad r_c^3=r_0^3\sinh\alpha\,\cosh\alpha
~,}
where $r_c$ is the charge radius of the black brane \refs{\CaldarelliXZ,\EmparanHG}.
Nevertheless, as in the DBI case \CallanKZ, and especially in extremal situations, 
the naive extrapolation of the leading order result beyond this regime continues to give qualitatively 
and quantitatively sensible results.

\subsec{1/4-BPS spike}

Only the $+$ branch in \statioci, \statiock\ has a sensible extremal limit. In this limit, where 
$T\to 0$ and $\beta \to +\infty$, the action \statioci\ simplifies to
\eqn\spikeaa{
I=\Omega_{(3)} L_t L_{x^1} Q_5
\int d\sigma \, \sigma^3 \sqrt{1+\frac{\kappa^2}{\sigma^6}}\, \sqrt{1+{z'}^2}\,
~.}
We are looking for a spike solution to the equations of motion of this action with the natural
boundary conditions
\eqn\spikeab{
\lim_{\sigma \to +\infty} z(\sigma)=0~, ~~
\lim_{\sigma \to 0^+} z'(\sigma)=-\infty
~.}
Such a solution exists and takes the simple analytic form
\eqn\spikeac{
z(\sigma)=\frac{|\kappa|}{2\sigma^2}
~.}
The validity of approximation \statioco\ breaks down when 
\eqn\spikead{
\sigma\sim r_c(\sigma) ~~ 
\Leftrightarrow ~~ \sigma \sim \sigma_c=\left(\frac{\pi N_5}{\sqrt{2}}\right)^{1/3}\left(1+\sqrt{1+\frac{64N_2^2}{N_5^4}}\ \right)^{1/6}\ell_P
~,}
where we made use of \statiobm,\statiocb.
Irrespective of this breakdown, we observe 
that the leading order solution is well-defined for all $\sigma\in \IR_+$ and, as we will see in a 
moment, the naive extrapolation beyond the strict regime of validity \statioco\ continues to give 
sensible results. We propose that the solution \spikeac\ captures the large-$(N_2, N_5)$ version of 
the 1/4-BPS M2-M5 intersection and the corresponding supersymmetric self-dual string soliton.

The energy density of the solution at the center of the soliton, at $\sigma=0$, corroborates this 
claim. The energy density can be evaluated from the on-shell value of the integrand of 
\spikeaa. A straightforward computation gives
\eqn\spikeae{
\frac{1}{L_t L_{x_1}} \frac{d I}{dz} \bigg |_{\sigma=0}=Q_2=N_2 T_{M2}
}
reproducing correctly the tension of $N_2$ BPS M2 branes. $T_{M2}$ denotes the tension of a
single M2 brane.

In addition, the transverse scalar profile \spikeac\ reproduces the $\frac{1}{\sigma^2}$ dependence
of the Howe-Lambert-West result \introab\ in the case of a single M5 brane. The only difference
lies in the scalar charge coefficient: $2Q_{sd}\sim N_2$ in the case of \HoweUE\ and 
$\frac{1}{4\pi^2} \frac{Q_2}{Q_5}\sim \frac{N_2}{N_5}$ in our case. This seems to imply that
the effective transverse scalar degree of freedom of the blackfold description is an average over the 
M5 branes, which is presumably a sign of the importance of abelian dynamics in the supersymmetric 
non-thermal case. A similar situation is encountered in supertubes \EmparanUX.
It would be useful to obtain a better understanding of the more general (holographic) relation 
between the blackfold effective degrees of freedom and the microscopic degrees of freedom of the 
multiple M5 brane theory.\foot{For transverse scalars the origin is common in both descriptions: they 
are Goldstone bosons associated with the breaking of translational symmetry. The abelian nature of 
the classical gravitational description emerges from the large-$N$ non-abelian nature of the multiple 
M5 brane theory.} Regardless of the specifics of this relation it is interesting to note the direct 
analogies between the way the known non-gravitational M5 brane worldvolume description works 
and how the blackfold description repackages the information of the gravitational solutions. This is 
one of the conceptual advantages of the blackfold approach.

\subsec{Thermal spikes}

By adding temperature to the above configuration the corresponding black brane
intersection becomes non-extremal. 
Following the general discussion of subsection 3.1 we are now looking for a spike solution 
of the $+$ branch equation \statiocm\ at finite $\beta$. However, unlike the zero-temperature
case such a solution does not exist over the full range of $\sigma$, $i.e.$ for 
$\sigma\in \IR_+$. The failure to obtain a sensible solution below a certain value of $\sigma$
is immediately obvious from eqs.\ \statiocf, \statiocg. For finite $\beta$, as we decrease $\sigma$
we reach a critical breakdown value, $\sigma_b$, where the term under one of the square roots 
becomes zero and then negative. This critical value equals
\eqn\thermalaa{
\sigma_b=\left( \frac{4 q_2^2}{\beta^6-4 q_5^2} \right)^{\frac{1}{6}}
~.}
The inequality \statiocd\ guarantees that the denominator in this expression is non-negative.

From the small temperature expansion of \thermalaa\ at fixed $q_2, q_5$,
\eqn\thermalab{
\sigma_b=\frac{(2|q_2|)^{\frac{1}{3}}}{\beta} \left( 1+\frac{2}{3} \frac{q_5^2}{\beta^6}+\ldots\right)
~,}
and the expression for the critical point of breakdown of the validity of the approximation for the 
extremal spike \spikead\ we deduce that up to leading order in temperature
\eqn\thermalac{
\frac{\sigma_b}{\sigma_c}=\frac{\sqrt{2}\ |\kappa|^{1/3}}{\beta} \left(1+\sqrt{1+\frac{64N_2^2}{N_5^4}}\ \right)^{-1/6} \ll 1
~.}
Hence, the breakdown of the leading order thermal spike solution occurs (at least within 
the small temperature expansion) well within the region where the leading order blackfold 
approximation cannot be trusted. In that sense, the pathological region is automatically
excised and poses no particular concern. The only issue we have to worry about is the issue
of boundary conditions. Which one of the solutions of the differential equation \statiocm\
does one pick for a given temperature? We will discuss the more general solutions of the 
differential equation \statiocm\ in the following subsection.

The same issue was encountered for thermal spikes of the F1-D3 system in \GrignaniMR. The 
strategy adopted in that paper was based on finding a matching point where an F1-D3 thermal spike
solution could be glued to a non-extremal black F-string at the desired temperature. More 
precisely, the thermal spike solution was chosen to reproduce the tension of a 
non-extremal black F-string. 

An analogous approach can be taken in our case. We can fix the solution of the differential
equation \statiocm\ by matching the tension of a planar black M2 brane to the local tension
\eqn\thermalad{
\frac{1}{L_t L_{x^1}}\frac{d M}{dz} \bigg |_{\sigma=\sigma_0}
}
of the thermal spike at a suitably chosen temperature-dependent $\sigma_0$. 
Since many of the details go in complete analogy with the BIon case of \GrignaniMR\ we will not 
discuss them further in this note. The resulting solution describes a thermalized self-dual string 
soliton solution.

\subsec{Wormhole solutions}

The most general solution of the differential equation \statiocm\ with the boundary condition
\eqn\wormaa{
\lim_{\sigma\to +\infty} z(\sigma)=0
}
is parametrized by a value $\sigma_0$, which is defined so that
\eqn\wormab{
\lim_{\sigma\to \sigma_0^+} z'(\sigma)=-\infty
~.}
Integrating the differential equation \statiocm\ with these boundary conditions we find
\eqn\wormac{
z_\pm(\sigma)=\int_\sigma^{+\infty}ds\, 
\left( \frac{F_\pm(s)^2}{F_\pm(\sigma_0)^2}-1\right)^{-\frac{1}{2}}
~.}

In this form the solution extends over the range $\sigma\in [\sigma_0,+\infty)$ and one has to 
decide how to extend it beyond this domain. In the previous subsection we considered the 
possibility of gluing a planar black M2 brane. Another possibility is to glue back at $\sigma_0$ 
the same solution with the opposite orientation. The resulting configuration describes a bi-funnel, or
wormhole-like solution that stretches between a stack of M5 and anti-M5 branes. Analogous
configurations for the BIon were considered in \refs{\CallanKZ,\GrignaniXM}. A configuration 
at non-zero $\sigma_0$ can be extremal but not BPS.

In what follows we summarize some of the main features of the wormhole solutions.
It will be convenient to define the distance between the M5 and anti-M5 stacks as
\eqn\wormad{
\Delta:=2\int_{\sigma_0}^{+\infty} ds\, \left( \frac{F(s)^2}{F(\sigma_0)^2}-1 \right)^{-\frac{1}{2}}
}
where $F=F_\pm$. Notice the scaling property
\eqn\wormae{
\Delta(\sigma_0;T,\kappa)=\kappa^{\frac{1}{3}}\, 
\Delta\left( \frac{\sigma_0}{\kappa^{\frac{1}{3}}};T,1\right)
~.}
A configuration exists only if $\sigma_0>\sigma_b$.

\ifig\loc{Plot of the distance $\Delta$ as a function of $\sigma_0$ for $\kappa=1$ in the extremal 
case.} {\epsfxsize3in\epsfbox{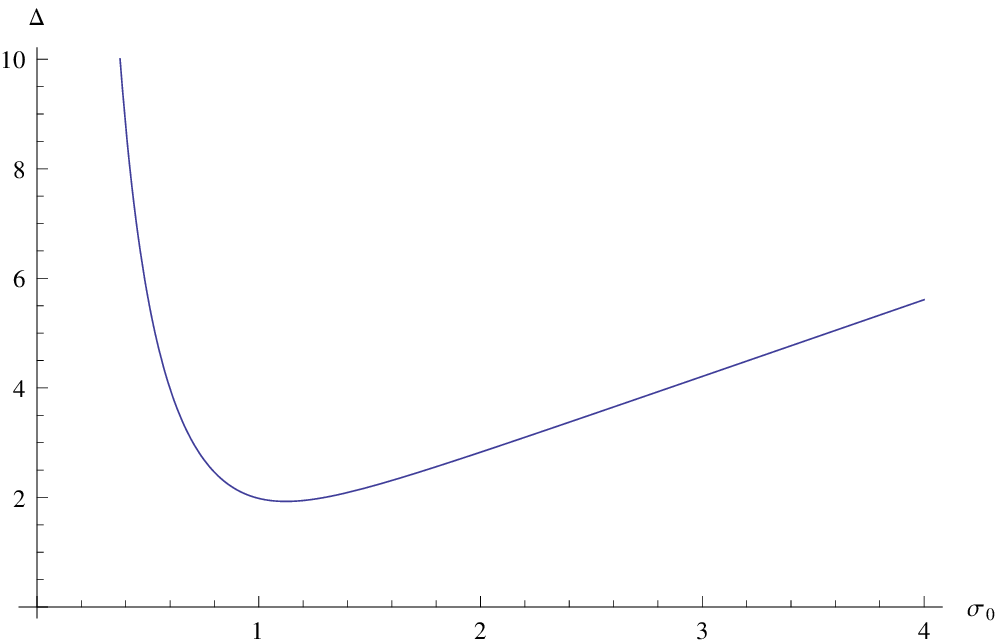}} 

\subsubsec{ \it Extremal wormholes}

Once again, only the $+$ branch is relevant for the extremal limit. The solution has an analytic
form for any $\sigma_0$
\eqn\wormba{
z(\sigma)=\int_\sigma^{+\infty} ds\, \sqrt{\frac{\sigma_0^6+\kappa^2}{s^6-\sigma_0^6}}
=\frac{\sqrt{\sigma_0^6+\kappa^2}}{2\sigma^2} \,  
_2 F_1\left( \frac{1}{3},\frac{1}{2},\frac{4}{3};\frac{\sigma_0^6}{\sigma^6}\right)
~.}
The corresponding distance $\Delta$ reads
\eqn\wormbb{
\Delta=\frac{\Gamma\left(\frac{1}{3}\right) \Gamma\left( \frac{1}{6} \right)}{6\sqrt{\pi}}
\frac{\sqrt{\sigma_0^6+\kappa^2}}{\sigma_0^2}
\simeq 1.402\, \frac{\sqrt{\sigma_0^6+\kappa^2}}{\sigma_0^2}
~.}
Its behavior as a function of $\sigma_0$ is depicted in Fig.\ 1. Analogous wormhole solutions 
can be found in the case of a single M5 brane using the fivebrane worldvolume theory 
\refs{\HoweFB\BandosUI-\AganagicZQ} or by uplifting to M-theory the BIon solutions of \CallanKZ.

We observe that there is a minimum distance
\eqn\wormbc{
\Delta_{\min}=\frac{\Gamma\left( \frac{1}{3} \right) \Gamma \left( \frac{1}{6} \right)}
{2^{\frac{4}{3}}\sqrt{3\pi}} \kappa^{\frac{1}{3}}
}
between the two fivebranes that occurs for $\sigma_{0,\min}=2^{\frac{1}{6}}\kappa^{\frac{1}{3}}$.
Hence, for a fixed distance $\Delta>\Delta_{\min}$ there are two possible solutions for $\sigma_0$.
In the large $\Delta$ limit they behave as
\eqn\wormbd{\eqalign{
({\rm thick~throat})~~~ &\sigma_0\simeq a \Delta  ~,
\cr
({\rm thin~throat})~~~ &\sigma_0\simeq \sqrt{\frac{\kappa}{a \Delta}}~, ~~
a:=\frac{6\sqrt{\pi}}{\Gamma\left( \frac{1}{3} \right) \Gamma \left( \frac{1}{6} \right)}\simeq 0.714
~.}}

\ifig\loc{Plots of $\Delta$ as a function of $\sigma_0$ for $\kappa=1$. The blue and red lines 
correspond to the non-extremal and extremal cases respectively. The left plot is done for 
$T=0.01$ and the right plot for $T=0.05$.}
{\epsfxsize2.6in\epsfbox{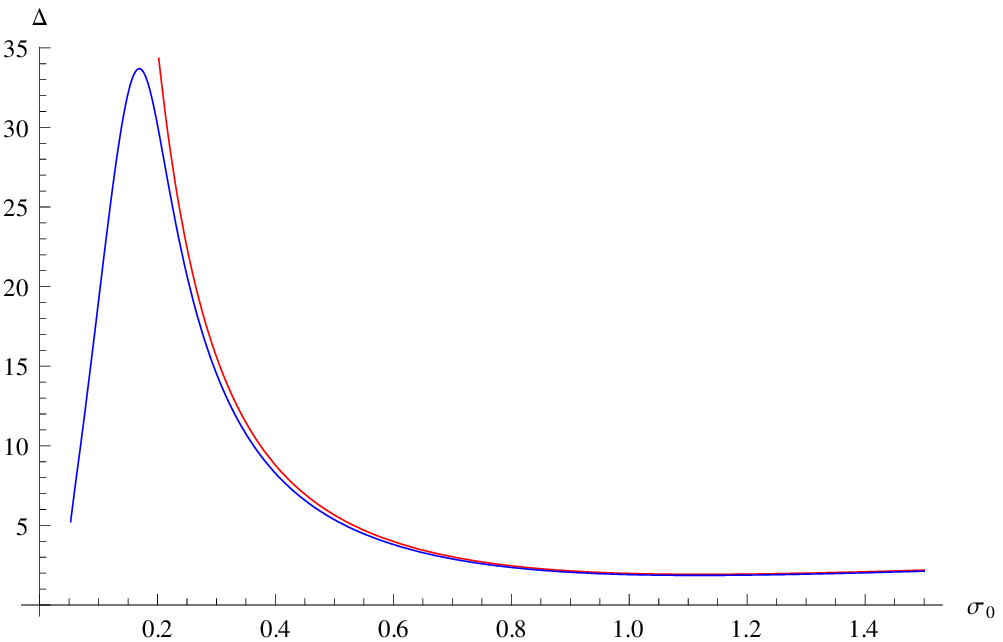}\epsfxsize2.6in\epsfbox{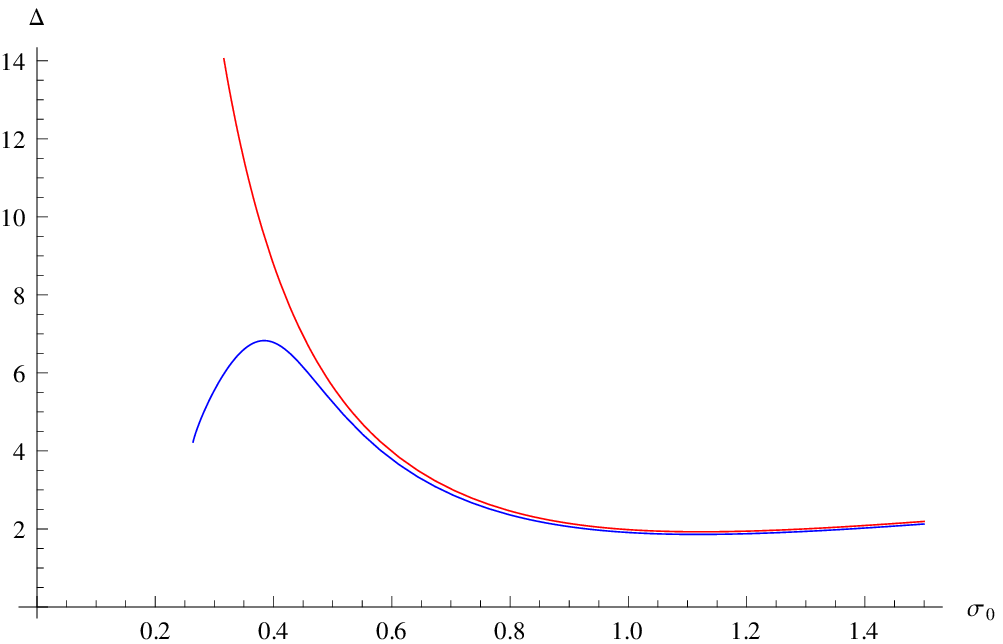}}

\subsubsec{ \it Branch connected to extremal wormholes}

For the $+$ branch and finite $\beta$ the distance $\Delta$ is given by the expression \wormad\
with $F=F_+$. We have not been able to find a closed analytic expression for generic values of
the temperature. The small temperature expansion takes the form
\eqn\wormca{
\Delta=\Delta_0+\beta^{-6} \Delta_1+\OO \left ( \beta^{-12} \right) 
}
where $\Delta_0$ is the extremal result \wormbb\ and 
\eqn\wormcb{
\Delta_1=-\frac{ \Gamma\left( \frac{1}{3} \right) \Gamma \left( \frac{1}{6} \right)}{180 \sqrt{\pi}}
\frac{2\kappa^2+5\sigma_0^6}{\sigma_0^{14}} \sqrt{\kappa^2+\sigma_0^6} ~ q_2^2 
~.}
Accordingly, the minimum we observed before is shifted to
\eqn\wormcc{
\sigma_{0,\min}=2^{\frac{1}{6}}\kappa^{\frac{1}{3}}
-\frac{7 q_5^2 \kappa^{\frac{1}{3}}}{2^{\frac{17}{6}}5} \beta^{-6}+\OO\left (\beta^{-12}\right)
}
and the corresponding minimum distance reads
\eqn\wormcd{
\Delta_{\min}=\frac{\Gamma\left( \frac{1}{3} \right) \Gamma \left( \frac{1}{6} \right)}
{2^{\frac{4}{3}}\sqrt{3\pi}} \kappa^{\frac{1}{3}}
\left( 1-\frac{q_5^2}{10\beta^6} \right)+\OO\left(\beta^{-12}\right)
~.}

\ifig\loc{Plots of $\Delta$ as a function of $\sigma_0$ for $\kappa=1$. The blue and red lines 
correspond again to the non-extremal and extremal cases respectively. The left plot is done
for $T=0.1$ and the right plot for $T=0.15$.}
{\epsfxsize2.6in\epsfbox{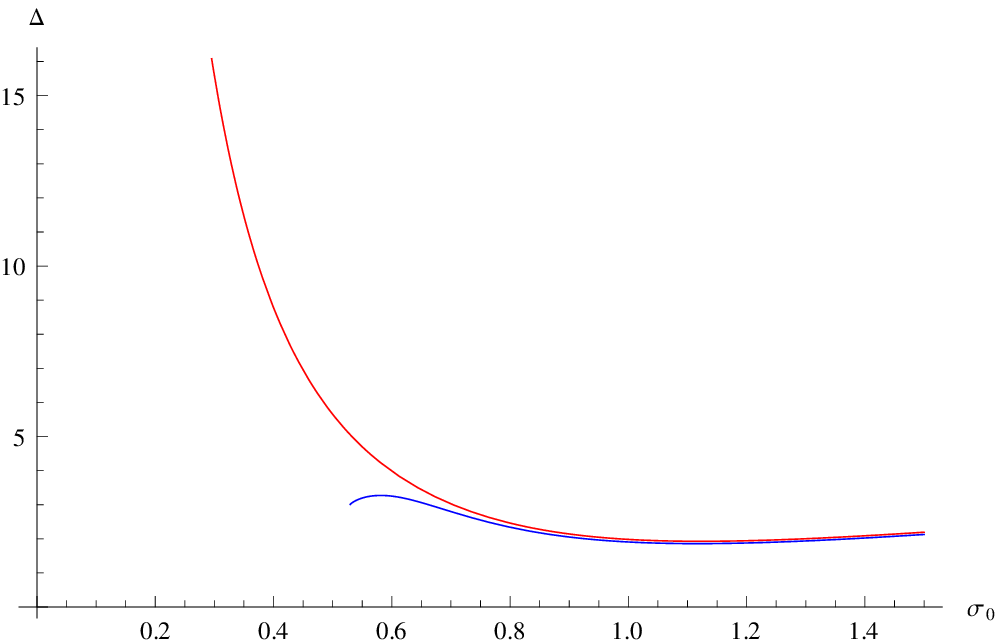}\epsfxsize2.6in\epsfbox{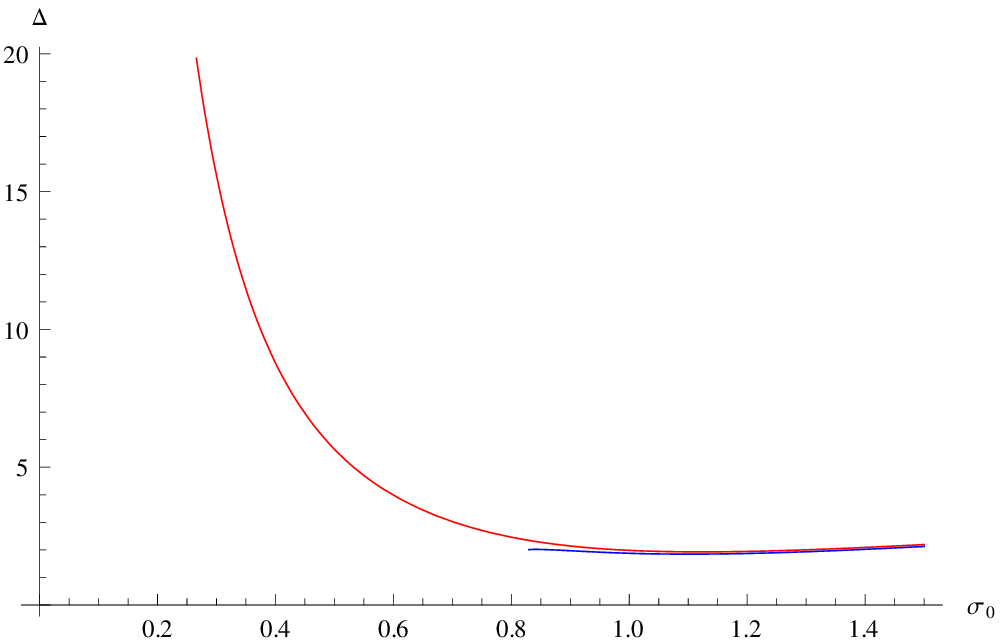}}

Plots of the distance as a function of $\sigma_0$ appear in Figs.\ 2, 3 for different
values of the temperature. A qualitatively new feature of the non-vanishing
temperature plots is the presence of a maximum. This maximum is already visible in the
perturbative expansion \wormca. At leading order
\eqn\wormce{\eqalign{
&\sigma_{0,\max}=\left( \frac{7}{12} \right)^{\frac{1}{12}} q_5^{\frac{1}{6}} \kappa^{\frac{1}{3}}
\beta^{-\frac{1}{2}}+\ldots~,
\cr
&\Delta_{\max}=\frac{17 \Gamma\left( \frac{1}{3}\right) \Gamma\left( \frac{1}{6} \right)}
{147\sqrt{\pi}} \left( \frac{15}{7} \right)^{\frac{1}{3}} q_5^{-\frac{1}{3}} \kappa^{\frac{1}{3}}\beta
+\ldots \simeq 1.254 ~q_5^{-\frac{1}{3}} \kappa^{\frac{1}{3}} \beta +\ldots
~.}}

Hence, we arrive at the following picture. At non-vanishing temperature the number of possible
solutions for a fixed distance $\Delta$ can vary from zero to three depending on the range of
$\Delta$. For $\Delta>\Delta_{\max}$ there is a single solution. For 
$\Delta_{\min}<\Delta<\Delta_{\max}$ there can be two or three solutions. In the vicinity of 
$\Delta_{\max}$ there are three solutions. As we lower $\Delta$ the solution with the lowest  value of 
$\sigma_0$ disappears if the critical value $\sigma_b$ \thermalaa\ is reached. For 
$\Delta<\Delta_{\min}$ the existence of a solution depends on whether a value of 
$\sigma_0>\sigma_b$ is possible. 

Analogous features have been observed for the thermal BIon solution \GrignaniMR.

\subsubsec{ \it Wormholes of the neutral branch}

The solution based on the $F_-$ function connects naturally to the 
neutral black fivebrane solution. This can be seen in the following manner.

In the low temperature limit the corresponding solution has the expansion 
\eqn\wormda{
z(\sigma)=\left(1+\frac{3q_2^2}{2\beta^6\sigma_0^6}\right)
\frac{\sigma_0^3}{2\sigma^2}\ {_2F_1}
\left(\frac{1}{3},\half,\frac{4}{3};\frac{\sigma_0^6}{\sigma^6}\right)+{\cal O}(\beta^{-12})
~.}
An analogous expression can be obtained by taking the neutral fivebrane limit
\eqn\wormdb{
q_5\to 0~, ~~ q_2={\rm finite}~, ~~ \hat \beta:=\frac{\beta}{\kappa^{\frac{1}{3}}}={\rm finite}
~.}
Expanding in powers of $q_5$ we find
\eqn\wormdc{
z(\sigma)=\left(1+\frac{3 q_5^2}{2\hat\beta^6\sigma_0^6}\right)
\frac{\sigma_0^3}{2\sigma^2}\ {_2F_1}\left(\frac{1}{3},\half,\frac{4}{3};\frac{\sigma_0^6}{\sigma^6}\right)+{\cal O}(q_5^4)
~}
that matches \wormda\ at leading order.

\newsec{Summary and further work}
\seclab\summary

Extremal and non-extremal black brane intersections are interesting supergravity solutions. In 
string/M-theory they contain useful information about the structure of the theory. Unfortunately, 
in many cases the generic complexity of these solutions does not permit to find fully localized 
supergravity intersections in closed analytic form. For that reason this is an opportune context
for the application of perturbative effective field theory descriptions like the blackfold approach.
Blackfolds provide a tractable and intuitive description of black brane dynamics by repackaging the 
gravitational information in an effective worldvolume formalism. The resulting expressions share 
similarities with the microscopic non-gravitational worldvolume descriptions of 
D-branes and M-branes in string/M-theory.

In this paper we have applied this formalism to the basic M2-M5 intersection 
extending previous work on the F1-D3 system \refs{\GrignaniXM,\GrignaniMR}. 
Our main purpose has been to demonstrate
how the formalism works in a simple representative situation and to relate the basic results  
with previous standard results in the literature of the M2-M5 system. In particular, we have seen
$(i)$ how one recovers the 1/4-BPS self-dual string soliton solution extending the single M5 brane 
result of \HoweUE\ to the regime of many M2 and M5 branes, and $(ii)$ how one can access the 
properties of the self-dual string soliton at finite temperature. The discussion of the supersymmetric
self-dual string soliton is directly related to the exact supergravity analysis of \LuninMJ. The 
non-extremal configurations in this paper provide, to the best of our knowledge, the first 
information for this type of black brane intersections in eleven dimensional supergravity.

The approach can be used to further probe the M2-M5 system in more generic situations. In a
companion paper \second\ we discuss M2-M5 intersections at finite temperature
and angular momentum. We present the corresponding solutions and compute their 
thermodynamic properties. 

In this note we have focused on the M2-M5 system in flat space. It is 
equally possible to discuss it in other backgrounds, for instance in AdS and within the context
of the AdS/CFT correspondence. Analogous discussions in AdS using the worldvolume description 
of a single M5 brane have appeared in \refs{\FayyazuddinQK\ChenIR-\ChenTT}. Blackfolds
in AdS have been analyzed in the past in \refs{\CaldarelliPZ\ArmasHZ-\GrignaniIW}.

Perhaps the most pressing question is whether we can use the approach presented in this work
to obtain new information about some of the currently inaccessible properties of the self-dual string 
soliton (and corresponding properties of the M5 brane). The fact that we can access the system 
in the regime of many M5 branes (which lies beyond the reach of most other methods) is 
encouraging. Since we work in the supergravity regime with an effective field theory tool the relation 
with the still illusive microscopic description of the M5 theory is indirect, however, it is not 
unreasonable to expect that the information obtained with our approach can provide new useful 
clues about the microscopic structure. Work in this direction is currently underway.

\bigskip
\centerline{\bf Acknowledgements}
\medskip

We would like to thank Anirban Basu, Niels Obers and Konstadinos Sfetsos for useful correspondence and 
discussions. The work of VN was partially supported by the European grants
FP7-REGPOT-2008-1: CreteHEPCosmo-228644, PERG07-GA-2010-268246, and
the EU program ``Thalis'' ESF/NSRF 2007-2013. KS has been supported by the ITN programme 
PITN-GA-2009-237920, the ERC Advanced Grant 226371, the IFCPAR CEFIPRA programme 
4104-2 and the ANR programme blanc NT09-573739. KS would like to thank the University 
of Patras and the University of Ioannina, for hospitality while part of this work was done.

\listrefs
\end